\pdfoutput=1
\documentclass[12pt,preprint]{aastex}
\usepackage{graphicx}
\usepackage{natbib}
\usepackage{aas_macros}
\usepackage[section] {placeins}
\usepackage{subfigure}
\usepackage{float}

\def\msun{{\rm\,M_\odot}}

\def\msun{{\rm\,M_\odot}}

\newcommand{\kms}{\, {\rm km\, s}^{-1}}

\newcommand{\mpc}{\, {\rm Mpc}}
\newcommand{\kpc}{\, {\rm kpc}}
\newcommand{\hmpc}{\, h^{-1} \mpc}

\newcommand{\hi}{\mbox{H\,{\scriptsize I}\ }}
\newcommand{\heii}{\mbox{He\,{\scriptsize II}\ }}

\def\h2{${\rm\,H_2}$}

\makeatletter 

\def\hi{\noindent \hangindent=2.5em}

\def\kpc{{\rm\,kpc}}

\def\mpc{{\rm\,Mpc}}

\def\kms{{\rm\,km/s}}
\def\msun{{\rm\,M_\odot}}

\def\vol#1  {{{#1}{\rm,}\ }}

\def\hi{{H\thinspace I}\ }

\def\heii{{He\thinspace II}\ }

\newcount\refno
\newcount\rfno
\def\eq{$^{\the\refno\ }$\advance\refno by 1}
\def\ad{\advance\rfno by 1}

\def\clock{\count0=\time \divide\count0 by 60
     \count1=\count0 \multiply\count1 by -60 \advance\count1 by \time
     \number\count0:\ifnum\count1<10{0\number\count1}\else\number\count1\fi}

\def\myputfigure#1#2#3#4#5%
{\hskip0.03\textwidth\vskip#5pt
\makebox[0pt]{\hskip#2in
\includegraphics[width=#3\textwidth]{#1}}\vskip#4pt\hfill}

\newcount\refno
\newcount\rfno
\def\eq{$^{\the\refno\ }$\advance\refno by 1}
\def\ad{\advance\rfno by 1}

\begin{document}

\title{Inconsequence of Galaxy Major Mergers in Driving Star Formation at $z>1$: Insights from Cosmological Simulations}
 
\author{
Renyue Cen$^{1}$
} 
 
\footnotetext[1]{Princeton University Observatory, Princeton, NJ 08544;
 cen@astro.princeton.edu}

\begin{abstract}

Utilizing a {\it high-resolution ($114h^{-1}${\rm pc})} adaptive mesh-refinement 
cosmological galaxy formation simulation of the standard cold dark matter model
with a {\it large} (2000-3000 galaxies with stellar mass greater than $10^9\msun$) {\it statistical} sample,
we examine the role of major mergers in driving star formation at $z>1$ in a cosmological setting,
after validating that some of the key properties of simulated galaxies are in reasonable agreement with 
observations, including luminosity functions, SF history, effective sizes and damped Lyman alpha systems.
We find that major mergers have a relatively modest effect on star formation,
in marked contrast to previous idealized merger simulations of disk galaxies that
show up to two orders of magnitude increase in star formation rate.
At $z=2.4-3.7$, major mergers tend to increase the specific star formation rate
by $10-25\%$ for galaxies in the entire stellar mass range $10^9-10^{12}\msun$ probed.
Their effect appears to increase with decreasing redshift, but is capped at $60\%$ at $z=1.4-2.4$. 
Two factors may account for this modest effect.
First, SFR of galaxies not in major mergers are much higher at $z>1$ than local disk galaxy counterparts.
Second, most galaxies at $z>1$ have small sizes and contain massive dense bulges, which
suppress the merger induced structural effects and gas inflow enhancement. 
Various other predictions are also made that will provide verifiable tests of the model.

\end{abstract}
 
 
\section{Introduction}

Simulations of major gas-rich disk galaxy mergers have provided 
quantitative insights to gas inflows 
and central starbursts under idealized conditions \citep[e.g.,][]{1996Barnes,1996Mihos,2006Hopkins}.
These simulations have laid the foundation of the theoretical framework for almost all contemporary 
mainstream interpretations of observed extreme starbursting galaxies,
namely, the ultraluminous infrared galaxies (ULIRGs),
as well as of the formation of supermassive black holes \citep[e.g.,][]{2005DiMatteo}.
This framework is appealing, because
almost all observed ULIRGs in the local universe 
either are directly seen merging or apparently show signs of mergers (at least some significant interactions)
\citep[e.g.,][]{1985Joseph,1988Sanders, 1997Duc, 1998Lutz} and at least some luminous quasars 
live in galaxies under strong interactions \citep[e.g.,][]{1997Bahcall}.
What is known but not sufficiently stressed in the relevant context is that the local universe 
is very different from the younger one at $z>1$ when star formation was much more intensive.
As an example, a typical Lyman Break Galaxy (LBG) several times less massive than our own Galaxy
has a star-formation rate (SFR) that is about ten times that of the Galaxy
\citep[e.g.,][]{2003Steidel}.
Moreover, minor mergers and close interactions between galaxies are expected to be much more
frequent at high redshift that, cumulatively, may have important effects.
Furthermore, there are significant structural differences between local galaxies and 
those at high redshift in that high redshift galaxies are more compact in size and 
the majority of massive quiescent galaxies that have been measured appear to have dense bulges
\citep[e.g.,][]{1997Lowenthal, 2005Daddi, 2006Trujillo, 2006bTrujillo, 
2007Toft,2007Longhetti,2008Buitrago, 2008Cimatti, 2009vanDokkum, 2009Cappellari,2011vandeSande}.
Therefore, our current physical interpretation of extreme galaxy events
that is obtained based on linking local observations with substantially idealized major galaxy merger simulations 
may not pertain to the high redshift universe in general.

In this work we examine theoretically, in a cosmological setting,
the role of major mergers in driving star formation in the redshift range $z>1$,
utilizing a large-scale high-resolution galaxy formation simulation.
At each redshift from $z=1.4$ to $z=3.7$ the simulation contains
$2000-3000$ galaxies with stellar mass greater than $10^9\msun$ 
resolved at better than $114h^{-1}$pc.
Detailed merger histories of galaxies are tracked and (binary) major mergers,
defined to be those of stellar mass ratios greater than $1/3$,
are examined in comparison to those that do not experience major mergers.
We find that for galaxies with SFR in the range $1-1000\msun$/yr
and the stellar mass range ${\rm M}_{\rm star}=10^9-10^{12}\msun$ examined,
major mergers, on average, yield a modest, fractional boost of $0-60\%$ in specific SFR;
we do not find two orders of magnitude increase in SFR found in previous merger simulations of disk galaxies
\citep[e.g.,][]{1996Mihos}.
We show that the properties of simulated galaxies are in reasonable 
agreement with observations and give a physical explanation of the results.
Additional predictions are provided to further test the model.
The outline of this paper is as follows.
In \S 2 we detail our simulation (\S 2.1) and galaxy catalogs (\S 2.2).
Results are presented in \S 3, followed by \S 4 that gives a physical explanation of the results.
Conclusions are given in \S 5.

\section{Simulations}\label{sec: sims}

\subsection{Hydrocode and Simulation Parameters}

We perform cosmological simulations with the adaptive mesh refinement (AMR) 
Eulerian hydro code, Enzo 
\citep[][]{1999bBryan, 2009Joung}.  
First we ran a low resolution simulation with a periodic box of $120~h^{-1}$Mpc on a side.
We identified a region centered on a cluster of mass of $\sim 2\times 10^{14}\msun$ at $z=0$.
We then resimulate with high resolution of the chosen region embedded
in the outer $120h^{-1}$Mpc box to properly take into account large-scale tidal field
and appropriate boundary conditions at the surface of the refined region.
This simulation box is the same region as the ``C" run in \citep[][]{2011cCen}.
The refined region for ``C" run has a size of $21\times 24\times 20h^{-3}$Mpc$^3$.
The initial condition in the refined region has a mean interparticle-separation of 
$58h^{-1}$kpc comoving, dark matter particle mass of $1.3\times 10^7h^{-1}\msun$.
The refined region is surrounded by three layers (each of $\sim 1h^{-1}$Mpc) of buffer zones with 
particle masses successively larger by a factor of $8$ for each layer, 
which then connects with
the outer root grid that has a dark matter particle mass $8^4$ times that in the refined region.

We choose the mesh refinement criterion such that the resolution is 
always better than $114h^{-1}$pc physical, corresponding to a maximum mesh refinement level of $13$ at $z=0$.
The simulation includes
a metagalactic UV background
\citep[][]{1996Haardt},  
and a model for shielding of UV radiation by neutral hydrogen 
\citep[][]{2005Cen}.
They also include metallicity-dependent radiative cooling 
\citep[][]{1995Cen}.
Our simulation also solves relevant gas chemistry
chains for molecular hydrogen formation \citep[][]{1997Abel},
molecular formation on dust grains \citep[][]{2009Joung}
and metal cooling extended down to $10~$K \citep[][]{1972Dalgarno}.
Star particles are created in cells that satisfy a set of criteria for
star formation proposed by \citet[][]{1992CenOstriker}.
Each star particle is tagged with its initial mass, creation time, and metallicity;
star particles typically have masses of $\sim$$10^6\msun$.

Supernova feedback from star formation is modeled following \citet[][]{2005Cen}.
Feedback energy and ejected metal-enriched mass are distributed into 
27 local gas cells centered at the star particle in question, 
weighted by the specific volume of each cell, which is to mimic the physical process of supernova
blastwave propagation that tends to channel energy, momentum and mass into the least dense regions
(with the least resistance and cooling).
We allow the entire feedback processes to be hydrodynamically coupled to surroundings
and subject to relevant physical processes, such as cooling and heating. 
The total amount of explosion kinetic energy from Type II supernovae
for an amount of star formed $M_{*}$
with a Chabrier IMF is $e_{SN} M_* c^2$ (where $c$ is the speed of light)
with  $e_{SN}=6.6\times 10^{-6}$.
Taking into account the contribution of prompt Type I supernovae,
we use $e_{SN}=1\times 10^{-5}$ in our simulation.
Observations of local starburst galaxies indicate
that nearly all of the star formation produced kinetic energy 
is used to power galactic superwinds \citep[e.g.,][]{2001Heckman}.
Supernova feedback is important primarily for regulating star formation
and for transporting energy and metals into the intergalactic medium.
The extremely inhomogeneous metal enrichment process
demands that both metals and energy (and momentum) are correctly modeled so that they
are transported in a physically sound (albeit still approximate 
at the current resolution) way.
The kinematic properties traced by unsaturated metal lines in DLAs are
extremely tough tests of the model, which is shown to agree well with observations \citep[][]{2010Cen}.
As we will show below, the properties of galaxies produced in the simulation 
resemble well observed galaxies, within the limitations of finite resolution.

We use the following cosmological parameters that are consistent with 
the WMAP7-normalized \citep[][]{2010Komatsu} LCDM model:
$\Omega_M=0.28$, $\Omega_b=0.046$, $\Omega_{\Lambda}=0.72$, $\sigma_8=0.82$,
$H_0=100 h {\rm km s}^{-1} {\rm Mpc}^{-1} = 70 {\rm km} s^{-1} {\rm Mpc}^{-1}$ and $n=0.96$.


\subsection{Simulated Galaxy Catalogs}

We identify galaxies in our high resolution simulation using the HOP algorithm 
\citep[][]{1999Eisenstein}, operated on the stellar particles, which is tested to be robust
and insensitive to specific choices of concerned parameters within reasonable ranges.
Satellites within a galaxy are clearly identified separately.
The luminosity of each stellar particle at each of the Sloan Digital Sky Survey (SDSS) five bands 
is computed using the GISSEL stellar synthesis code \citep[][]{Bruzual03}, 
by supplying the formation time, metallicity and stellar particle mass.
Collecting luminosity and other quantities of member stellar particles, gas cells and dark matter 
particles yields
the following physical parameters for each galaxy:
position, velocity, total mass, stellar mass, gas mass, 
mean formation time, 
mean stellar metallicity, mean gas metallicity,
star formation rate,
luminosities in five SDSS bands (and various colors) and others.

We create catalogs of galaxies from $z=1.4$ to $z=3.7$ with an increment of $\Delta z=0.05$.
We track the merger history of each galaxy in this redshift span.
There are two different ways to define major mergers.
First, a theoretical one where we identify the merger time as that when two galaxies with a stellar mass ratio greater than $1/3$
are fully integrated into one
with no identifiable separate stellar peaks.
Second, an observational one
where a major merger is defined to be that where a galaxy has 
a neighbor galaxy with a stellar mass greater than $1/3$ its mass at a lateral distance smaller than $40$kpc proper.
Both will be used in subsequent analysis.
It is useful to state that the observationally-oriented definition does not
always lead to a true merger of the usual sense, because either the two galaxies are a projected pair,
or their merging time scale is much longer than the relevant dynamic time 
or the time before something else will have happened to the two concerned galaxies.
Some informative comparisons or distinctions between the two will be made, when useful.
We find that there are about 2000-3000 galaxies with stellar mass greater than $10^9\msun$ 
maximally resolved at better than $114h^{-1}$pc
at each redshift snapshot in the range $z=1.4-3.7$, providing us with unprecedented statistical power.

In \citet[][]{2011cCen} we show that galaxy luminosity functions for both UV and FIR selected galaxies 
can be self-consistently produced by the simulation.
This, in combination with other, independent tests of the simulation, including the properties 
of the damped Lyman alpha systems \citep[][]{2010Cen},
strongly indicates a range of applicability of our simulation to complex systems, including galaxies at
sub-kpc ISM scales.
This validation of the simulation results is critical and allows us, with significant confidence,
to perform the particular analysis here with respect to effects of major mergers.

\section{Results}

\begin{figure}[ht]
\centering
\vskip -0.0cm
\resizebox{5.0in}{!}{\includegraphics[angle=0]{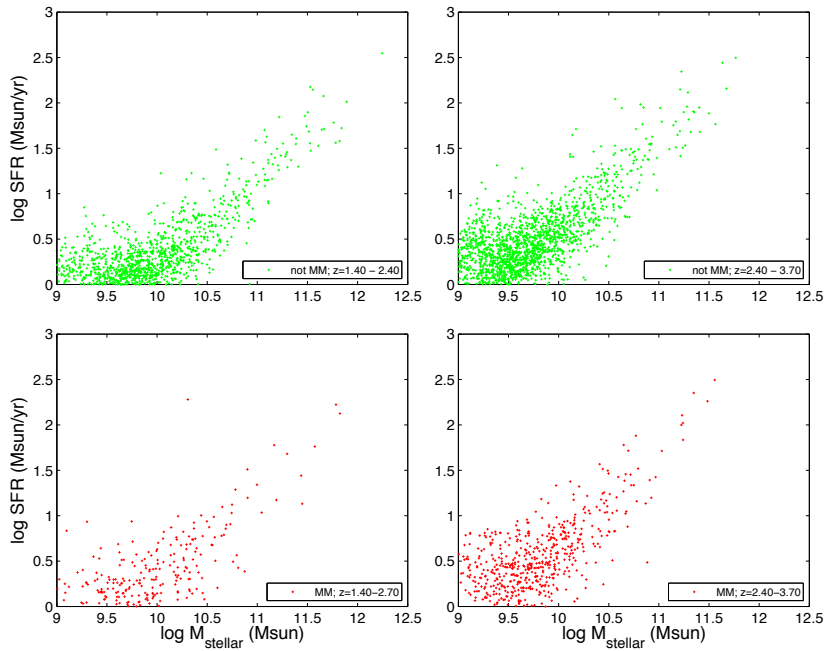}}
\vskip -0.6cm
\caption{\footnotesize 
places each galaxy as a plus symbol in the SFR-stellar mass plane
for non major merger galaxies in the redshift range $z=1.4-2.4$ (top left panel)
and $z=2.4-3.7$ (top right panel).
The corresponding ones for galaxies with major mergers are shown in the bottom panels.
Here we adopt the observationally oriented definition of major mergers, i.e.,
pairs of stellar mass ratio greater than $1/3$ and projected separation less than $40$kpc.
Only a small percentage of randomly selected galaxies is shown.
}
\label{fig:ulirg}
\end{figure}

Figure~\ref{fig:ulirg} shows scatter plots between SFR and stellar mass 
for galaxies that do not have ongoing major mergers (top two panels),
compared to those that are ongoing major mergers (bottom two panels).
Under visual inspection we see that there is no major discernible difference
between galaxies that do and do not experience major mergers
in the redshift range examined for the entire range of stellar mass or SFR.
It is noticeable that the number of galaxies that are major mergers
is a minor fraction of all galaxies at any stellar mass or SFR.

\begin{figure}[ht]
\centering
\vskip -0cm
\resizebox{5.0in}{!}{\includegraphics[angle=0]{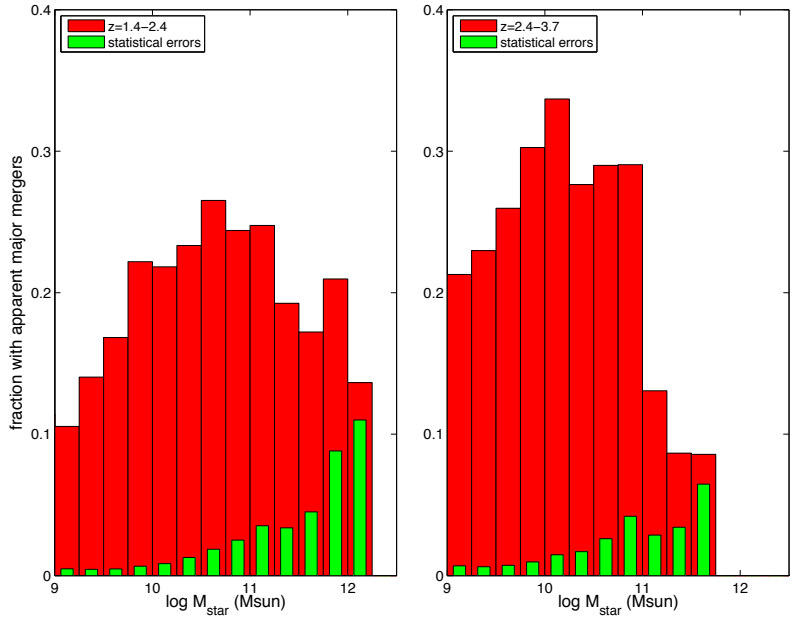}}
\vskip -0.6cm
\caption{\footnotesize 
shows the fraction of galaxies that are in major merger as a function of stellar mass (red histograms)
at $z=1.4-2.4$ (left panel) and $z=2.4-3.7$ (right panel).
The statistical errors are shown as green histograms.
We use the observationally oriented definition of major mergers, i.e.,
pairs of stellar mass ratio greater than $1/3$ and projected separation less than $40$kpc.
}
\label{fig:majorfrac}
\end{figure}

Figure~\ref{fig:majorfrac} shows the fraction of galaxies 
that are in major merger as a function of stellar mass with the observational definition.
We note that the major merger fraction at the low steller mass ($<10^{11}\msun$) 
is substantially overestimated due to the adopted definition,
because many satellite galaxies within the virial radius of large galaxies
are ``mis-identified" as major mergers in this case.
In fact, many of these satellite galaxies do not ever merge with one another directly in a binary fashion,
as will be shown below in Figure~\ref{fig:rate}.
The fraction of major mergers at the high stellar mass end 
does not significantly suffer from this ``projection" effect.
We see that for galaxies with stellar mass in the range $10^{11}-10^{12}\msun$
major merger galaxies make up about {\it $10-20\%$ of all galaxies} in that mass range.

\begin{figure}[ht]
\centering
\vskip -1cm
\resizebox{5.0in}{!}{\includegraphics[angle=0]{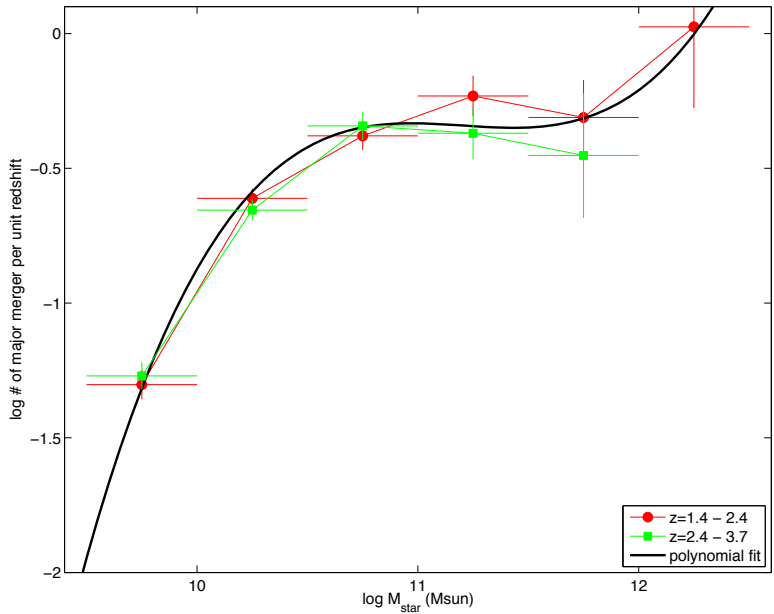}}
\vskip -1cm
\caption{\footnotesize 
shows the merger rate ($=$number of major mergers per unit redshift)
as a function of galaxy stellar mass for galaxies at $z=1.4$ (red dots) and $z=2.4$ (green squares).
Here a merger is more physically based definition, 
an event where two galaxies of the stellar mass ratio greater than $1/3$ physically merge.
}
\label{fig:rate}
\end{figure}

The results on major merger fractions shown in 
Figure~\ref{fig:majorfrac} (and Figure~\ref{fig:majorSFRfrac} below) are based on the observational
definition of major mergers.
It is useful to distinguish that from the theoretical one,
where the latter is based on the actual merger events rather than pairs within some projected distance.
Figure~\ref{fig:rate} shows the theoretical merger rate, defined to be the number of major mergers per unit redshift,
as a function of galaxy stellar mass for galaxies at $z=1.4$ (red dots) and $z=2.4$ (green squares),
respectively.
We see that the actual merger rate is roughly constant at $\sim 0.3-0.5$ per unit redshift 
for the stellar mass range ${\rm M}_{\rm star}=10^{10.5}-10^{12}\msun$.
At ${\rm M}_{\rm star}>10^{12}\msun$ there is hint for a significant upturn in merger rate,
albeit with less statistical certainty due to a small number of such massive galaxies in the simulation. 
Nevertheless, such an upturn would be 
consistent with the expectation that the central cD galaxies 
may experience more major mergers due to dynamical inspiral of satellites. 
This is also consistent with the apparent difference seen in 
Figure~\ref{fig:rate} between galaxies at $z=1.4-2.4$ (red)
and galaxies at $z=2.4-3.7$ (green) in that the upturn is absent 
in the higher redshift range, because of the absence of large galaxies at that redshift range in the given simulation box.
If the simulation box were large enough to contain cD-like galaxies at that higher-redshift range,
we expect to see the same upturn.

The downturn at ${\rm M}_{\rm star}<10^{10.5}\msun$ of the merger rate
is still more dramatic.
We see a decrease of merger rate by a factor of $\sim 10$ from ${\rm M}_{\rm star}=10^{10.5}\msun$ to 
${\rm M}_{\rm star}=10^{9.5}\msun$.
This should be compared to about a factor of $1.2-1.7$ drop seen 
in Figure~\ref{fig:majorfrac} across the mass range. 
This shows that the vast majority 
of galaxies of mass ${\rm M}_{\rm star}\le 10^{10}\msun$ 
that are seen in close proximity ($<40$~kpc) with other galaxies of comparable masses
are in fact do not end up in binary major mergers.
In \citet[][]{2011cCen} we show that the simulation reproduces observed luminosity functions
in the concerned redshift range, indicating that the simulation is ``complete" down to 
about a galaxy stellar mass of $\sim 10^9\msun$.
Thus, the results for the range of galaxy stellar mass shown here is reliable.
A plausible physical explanation for the sharp downturn at the low mass end may be that
most satellite galaxies just zoom around and never merge with their fellow satellite galaxies 
rather they dynamically spiral in to merge with the primary galaxy or remain as satellites.
A more detailed study focused on the demographics of mass accretion, including mergers,
will be presented elsewhere. 
Here we present a third-order polynomial fit to  the major merger rate, $R$, defined to be the number of
major mergers per unit redshift: 
\begin{equation}
\label{eq:R}
\log {\rm R}  = 0.34 (\log {\rm M}_{\rm star}-11)^3 - 0.21 (\log {\rm M}_{\rm star}-11)^2 - 0.013 (\log {\rm M}_{\rm star}-11) - 0.33,
\end{equation}
\noindent
shown as the solid black curve in Figure~\ref{fig:rate},
where ${\rm M}_{\rm star}$ is in solar masses.

\begin{figure}[ht]
\centering
\vskip -1cm
\resizebox{5.0in}{!}{\includegraphics[angle=0]{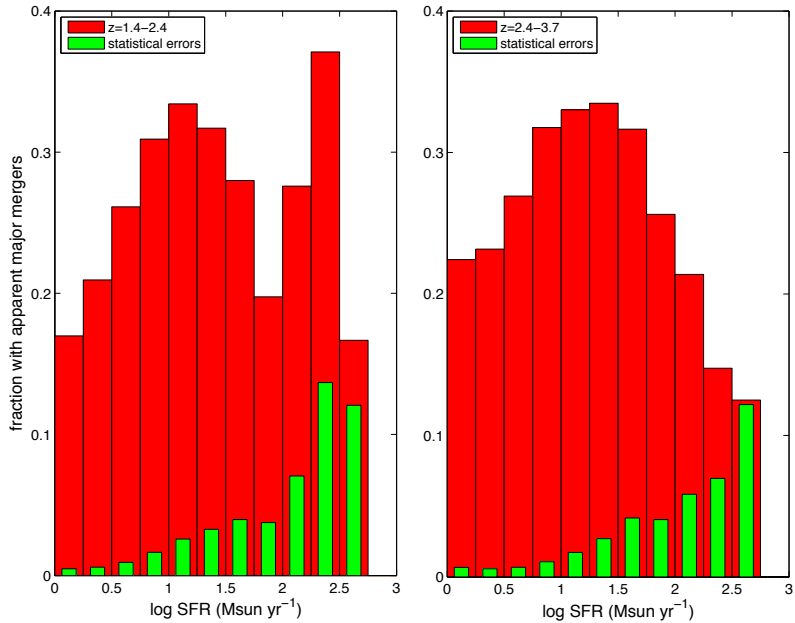}}
\vskip -1cm
\caption{\footnotesize 
shows the fraction of galaxies that are in major merger as a function of SFR (red histograms)
at $z=1.4-2.4$ (left panel) and $z=2.4-3.7$ (right panel).
The statistical errors are shown as green histograms.
We use the observationally oriented definition of major mergers, i.e.,
pairs of stellar mass ratio greater than $1/3$ and projected separation less than $40$kpc.
}
\label{fig:majorSFRfrac}
\end{figure}

Figure~\ref{fig:majorSFRfrac} shows the fraction of galaxies 
that are in major merger as a function of SFR.
Similar to Figure~\ref{fig:majorfrac}, the actual major merger fraction at the low SFR end shown is overestimated,
given the observational definition used.
The major merger rate at the high SFR end, at SFR$\ge 200\msun$~yr$^{-1}$, is less affected and
the simulation shows that one should expect to see $10-40\%$ of these high SFR galaxies to be in 
apparent major mergers.
This fraction is consistent with the observed upper bound of 
57\% (8/14) for the submillimeter galaxy (SMGs) sample of \citet[][]{2006Tacconi} at $z=2-3.4$
that show a double-peaked profile in the CO 3-2/4-3 emission.
Of this observed fraction of SMGs in major mergers, 
a fraction of it may be due to orbital motion of emitting gas in a disk configuration
or some other configurations instead of major mergers.
We predict that, when high spatial resolution become available with the upcoming ALMA mission,
{\it the fraction due to major mergers should be in the range $10-40\%$}, if our model is correct.
For star-forming galaxies of SFR$\le 200\msun$~yr$^{-1}$ at $z=1.4-3.7$,
we also predict that the major merger fraction should fall in the range of $15-35\%$.
Recall that here we use the observationally oriented definition of major mergers, i.e.,
pairs of stellar mass ratio greater than $1/3$ and projected separation less than $40$kpc.

\begin{figure}[ht]
\centering
\vskip -1cm
\resizebox{5.0in}{!}{\includegraphics[angle=0]{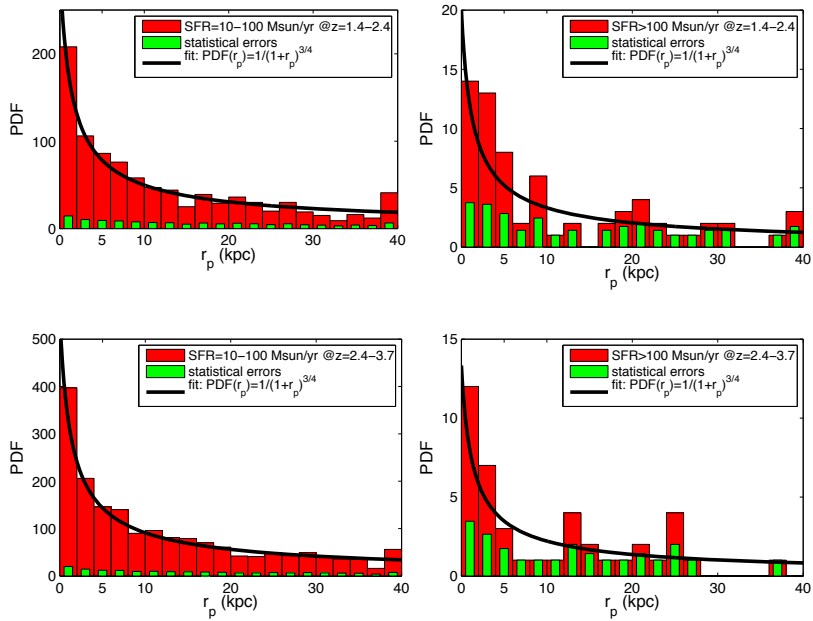}}
\vskip -1cm
\caption{\footnotesize 
shows the probability distribution functions (PDF) of the projected separation (${\rm r}_{\rm p}$)
of major mergers at $z=1.4-2.4$ (two upper panels) and $z=2.4-3.7$ (two bottom panels), respectively.
The left panels are for star-forming galaxies of SFR$=10-100\msun$~yr$^{-1}$ and
the right panels for star-forming galaxies of SFR$>100\msun$~yr$^{-1}$.
The red histograms are the PDFs and the green histograms the statistical errors at each bin.
The black curves show a power fit described by Eq~\ref{eq:sep}.
}
\label{fig:sep}
\end{figure}

To provide further tests of our model predictions,
Figure~\ref{fig:sep} shows the probability distribution functions (PDFs) of the projected separation (${\rm r}_{\rm p}$)
of major mergers at $z=1.4-2.4$ (top panels) and $z=2.4-3.7$ (bottom panels)
for two subgroups of galaxies of SFR$=10-100\msun$~yr$^{-1}$ (left panels) and SFR$>100\msun$~yr$^{-1}$ (right panels),
respectively.
We find that the PDFs are reasonably fit with a single cored powerlaw of the following form:
\begin{equation}
\label{eq:sep}
{\rm PDF}({\rm r}_{\rm p})d {\rm r}_{\rm p} \propto ({\rm r}_{\rm c}+{\rm r}_{\rm p})^{-3/4} d {\rm r}_{\rm p},
\end{equation}
\noindent
where the projected separation ${\rm r}_{\rm p}$ and core size ${\rm r}_{\rm c}$ are in physical kpc.
The black curves shown in Figure~\ref{fig:sep} have ${\rm r}_{\rm c}=1$, although 
it is not stringently constrained.
The simple powerlaw fits are quite good, in contrast to gaussian or exponential forms that
are found to provide poor fits. 
The found slope of $-3/4$ in the PDF suggests that the three-dimensional distribution
around each star-forming galaxy of other galaxies of comparable SFR 
approximately follows a powerlaw of a slope of $-2.75$.
Details of this and other related clustering issues of galaxies will be presented elsewhere.

\begin{figure}[ht] 
\centering
\vskip -1cm
\resizebox{5.0in}{!}{\includegraphics[angle=0]{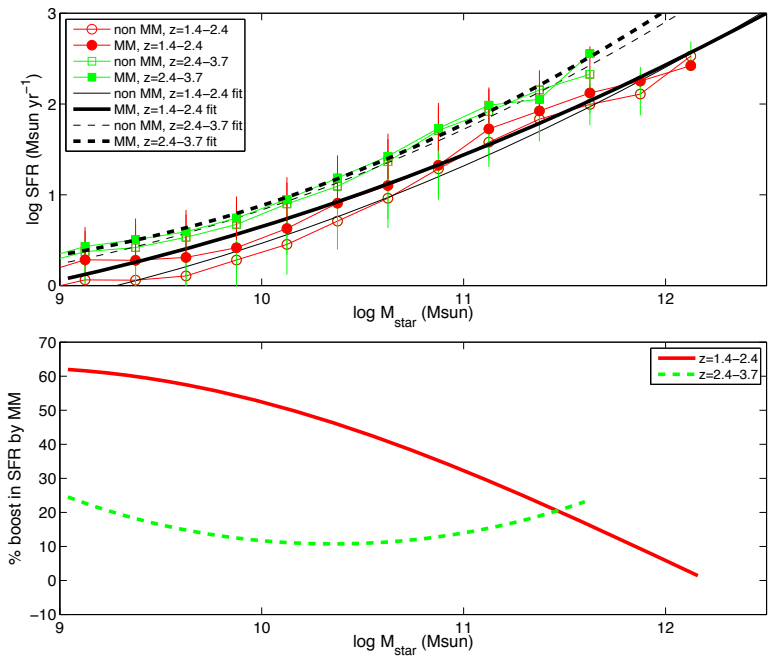}}
\vskip -1cm
\caption{\footnotesize 
Top panel: the mean SFR of galaxies at a given stellar mass
for galaxies that are in major mergers (red solid dots) and not in major mergers (red open dots) at $z=1.4-2.4$.
The corresponding ones at $z=2.4-3.7$ are shown in green squares.
The errorbars show the dispersion around the mean.
The thin and thick dashed curves are the best second-order polynomial fits to 
the non major mergers and major mergers, respectively, at $z=1.4-2.4$.
The thin and thick solid curves are the best second-order polynomial fits to 
the non major mergers and major mergers, respectively, at $z=2.4-3.7$.
We use the observationally oriented definition of major mergers, i.e.,
pairs of stellar mass ratio greater than $1/3$ and separation less than $40$kpc.
Bottom panel: the ratio of fitted curves to the major merger and non-major-merger minus one 
for $z=1.4-2.4$ (red solid curve) and $z=2.4-3.7$ (green dashed curve), respectively.
Visually the ratio of the fitted curves and the actual computed data points display comparable amplitudes.
}
\label{fig:majorSFR}
\end{figure}

\begin{figure}[ht]
\centering
\vskip -1cm
\resizebox{4.5in}{!}{\includegraphics[angle=0]{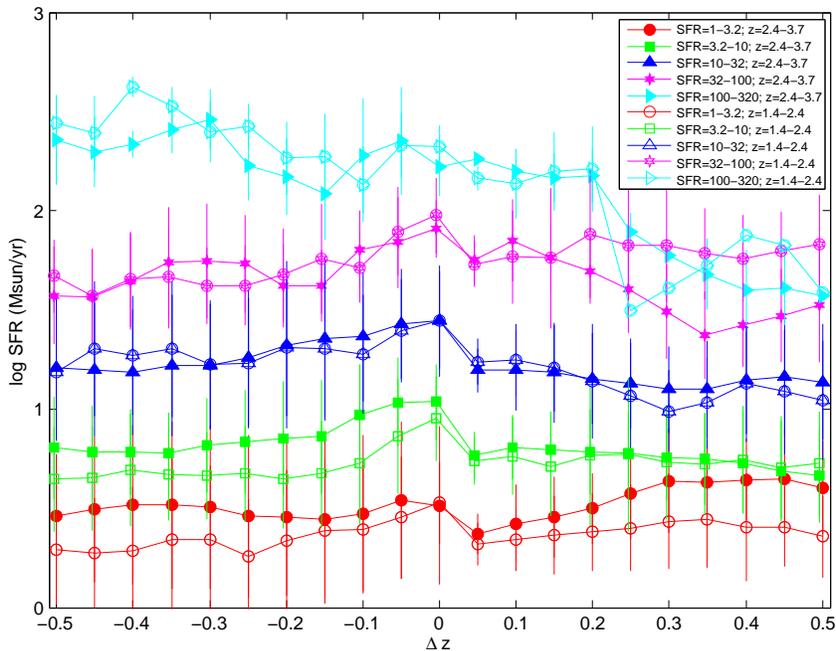}}
\vskip -1cm
\caption{\footnotesize 
shows the history of the mean SFR as a function of time redshift $\Delta z$ 
for five different subsets of galaxies with SFR at $\Delta z=0.05$ (i.e., prior to the merger event)
equal to $1-3.2$, $3.2-10$, $10-32$ and $32-100$, $100-320\msun$~yr$^{-1}$, respectively,
separately for galaxies in the redshift range $z=1.4-2.4$ and $z=2.4-3.7$.
Dispersions on the means are shown as well.
}
\label{fig:major}
\end{figure}

The top panel of Figure~\ref{fig:majorSFR} 
shows the meean SFR as a function of galaxy stellar mass, separately, for galaxies
that are in major mergers and galaxies that are not in major mergers.
The ratio of fitted curves for the galaxies with major mergers and those without major mergers minus one 
are shown in the bottom panel for $z=1.4-2.4$ (red solid curve) and $z=2.4-3.7$ (green dashed curve).
We see that major mergers appear to experience very modest boost in SFR for galaxies at $z=2.4-3.7$,
at about $10-25\%$ for the entire stellar mass range ${\rm M}_{\rm star}=10^9-10^{12}\msun$ probed.
The overall strength of the boost due to major mergers appear to increase with decreasing redshift,
when one compares the values at $z=1.4-2.4$ to those at $z=2.4-3.7$,
but remains at less than $60\%$ across the entire mass range.
It also appears that there may be a trend of a relatively larger boost of SFR due to major mergers
for lower mass galaxies than for larger mass galaxies at $z=1.4-2.4$.
But we caution that the results in the bottom panel are somewhat sensitive to the exact fits;
given that the fits do not exactly reproduce all the data points, one should 
be careful to not take the exact curves of the fits too literally.
In any case, it is abundantly clear that we {\it do not see 
very large increase in SFR of a factor of two orders of magnitude}
that are found in simulations of isolated major gas-rich spiral galaxy mergers \citep[e.g.,][]{1996Mihos}.

In Figure~\ref{fig:majorSFR} the modest boost in SFR due to major mergers is computed 
using the observationally oriented definition of major mergers, i.e.,
pairs of stellar mass ratio greater than $1/3$ and projected separation less than $40$kpc.
We now compute a similar quantity using 
the theoretical definition of major mergers where we identify the merger
time as that when two galaxies are fully integrated into one
with no identifiable separate stellar peaks.
We follow the history of each galaxy and ``stack" all major merger events 
centered at $\Delta z=0$.
Figure~\ref{fig:major} shows the mean SFR history for galaxies at 
five given ranges of SFR, measured at $\Delta z=0.05$ (using a different redshift, say, $\Delta z=0.10$ or 0.15, makes no 
material difference in the results).
In a fashion that is consistent with the findings shown in Figure~\ref{fig:majorSFR},
we do not find any dramatic boost of SFR at the merger redshift 
and at $|\Delta z| \le 0.5$ for galaxies at SFR$\le 100\msun$yr$^{-1}$ in the redshift range $z=1.4-3.7$.
The $1\sigma$ dispersion about the mean is about $1.5-3$, roughly consistent with 
the range of SFR for each subset at $\Delta z=0.05$, with a tendency that
the dispersion is larger for lower SFR subsets.
For the subset with the largest SFR ($\ge 100\msun$~yr$^{-1}$), however,
there is a visually noticeable jump in SFR by a factor of $\sim 2-5$
from $\Delta z > 0.2$ to $\Delta z < 0.2$, hinting an intriguing possibility 
that a major merger event, not necessarily the final major merger moment,
serves to ``trigger" a very high SFR event.
In other words, it suggests that some very high SFR galaxies, such as ULIRGs or SMGs,
may be {\it initially} triggered by some major merger events. 
At the same time results in Figure~\ref{fig:major} 
also suggest that the very high SFR ($\ge 100\msun$~yr$^{-1}$)
galaxies remain at the elevated and upward trend for SFR following the merger event 
for a long period of time ($\Delta z\sim 1$) that is much longer than the typical merger time scale.
This has profound implications for the nature of ULIRGs and SMGs that will be addressed elsewhere.

\section{Physical Explanation of the Results}

Both external gravitational and internal gravitational and hydrodynamic torques may drive gas inward.
Externally, the tidal field from a companion during a galaxy merger, major or minor,  
gives rise to a non-axisymmetric gravitational potential. 
This induces a response of the disk material \citep[][]{1972Toomre}, 
in particular its cold gas, stronger for prograde mergers.
More broadly, tidal fields from interacting galaxies, which are not necessarily merging with one another,
may drive gas inward.
Internally, non-axisymmetric gravitational potentials, notably those sustained by stellar bars that are produced by 
secular evolution of sufficiently cold stellar disks under certain conditions or from other interactions, such as mergers,
can also drive gas inward.
A thorough study of torques due to gravitational and hydrodynamic processes
to isolate the primary physical mechanisms governing 
the gas inflows in a cosmological setting will be performed in a larger study. 
Here we provide some physical insight for the results found, relying mainly on circumstantial but strong evidence.

\begin{figure}[ht]
\centering
\vskip -0.5cm
\resizebox{5in}{!}{\includegraphics[angle=0]{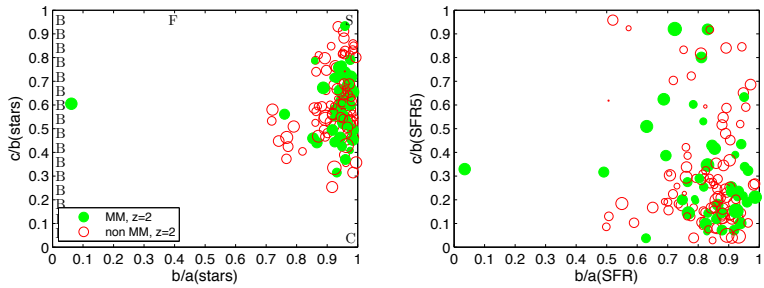}}
\vskip -5.0cm
\caption{\footnotesize 
Left panel: shows axial ratios c/b versus b/a, 
where $c<b<a$ are the semi-axes of an ellipsoid approximating the stellar distribution within ${\rm r_e}$
for galaxies with ${\rm SFR}\ge 10\msun$yr$^{-1}$
with major mergers (solid dots) and those without (open circles) at $z=2$.
The symbol size in both panels is linearly proportional to the logrithm of its SFR.
Several special locations are indicated by special letters:
``B" for thin bars of various thickness,
``S" for sphere,
``C" for flat circular disk
and ``F" for American football.
Right panel: shows the same but for SFR density distribution within the radius of 50\% SFR.
}
\label{fig:DeDe}
\end{figure}

Anecdotal evidence and visual examination of some galaxies 
suggest that chaotic gas inflows often result in mis-alignments of newly formed stellar disks with
previous stellar disk/non-spherical bulges, 
and the orbital planes of infalling satellite stellar or gas clumps do not always have 
a fixed orientation.
These processes cumulatively may be thought to create stellar distributions
in the central regions that are dynamically hot, which, in turn,
provides conditions that are unfavorable to secular formation of stellar bars.
We check if this indeed is the case.

\begin{figure}[ht]
\centering
\vskip -0.5cm
\resizebox{5in}{!}{\includegraphics[angle=0]{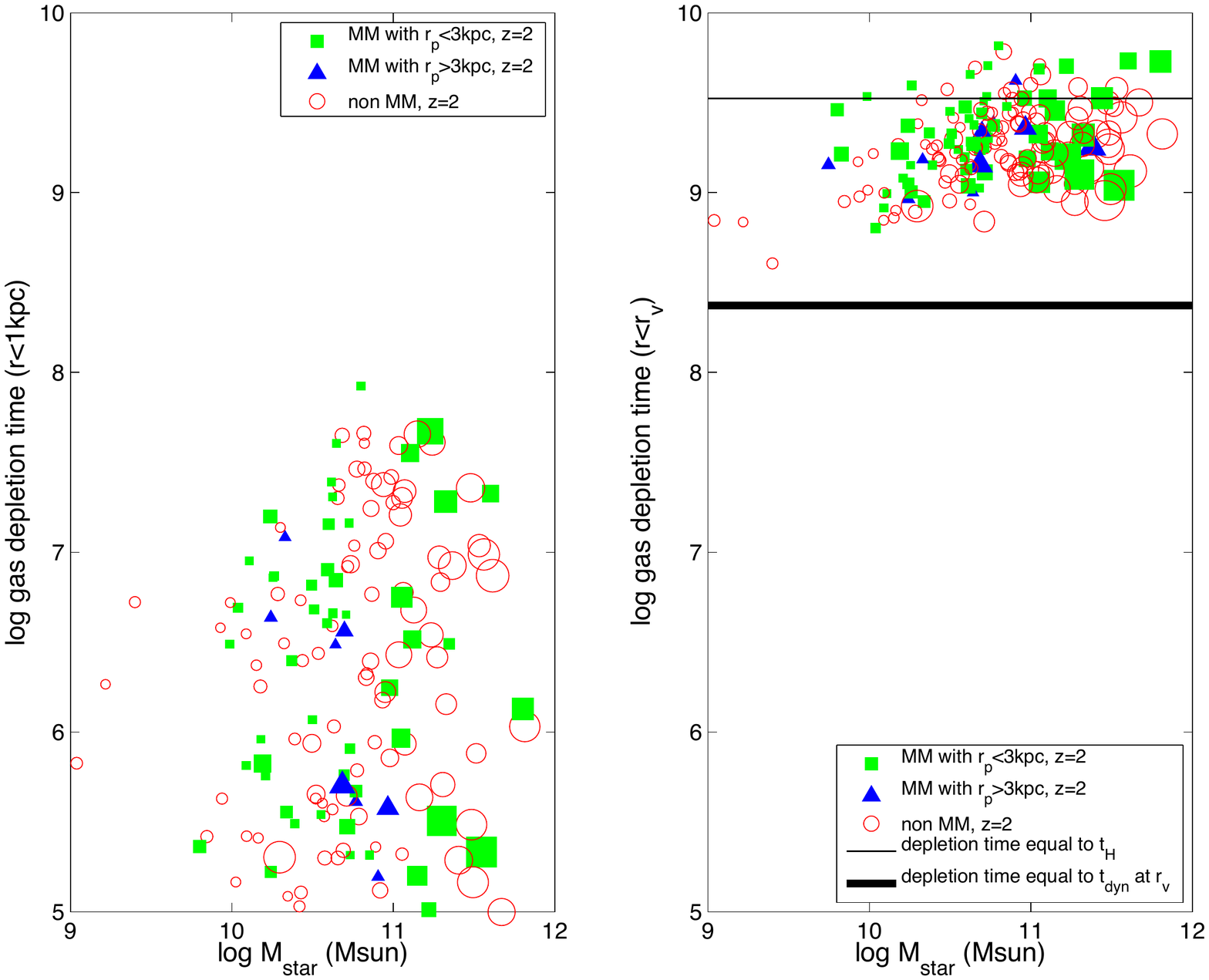}}
\vskip -0.5cm
\caption{\footnotesize 
Left panel: gas depletion time within the central $1$kpc region of galaxies at $z=2$ with 
${\rm SFR}\ge 10\msun$yr$^{-1}$.
Three types of galaxies are shown using different symbols: galaxies that are not undergoing major mergers are 
open circles, galaxies in major mergers with projected separation between the two galaxies
less than $3$kpc as squares and greater than $3$kpc as triangles.
Right panel: gas depletion time over the entire galaxy within the virial radius. 
The symbol size is linearly proportional to the logrithm of SFR.
The thin and thick horizontal lines correspond to the Hubble time and 
dynamical time at virial radius at $z=2$, respectively.
Here the observational definition of major mergers is used.
}
\label{fig:tgasdepletion}
\end{figure}

The left panel of Figure~\ref{fig:DeDe} shows axial ratios c/b versus b/a, 
where $c<b<a$ are the semi-axes of an ellipsoid approximating the stellar distribution within ${\rm r_e}$
for galaxies with ${\rm SFR}\ge 10\msun$yr$^{-1}$
with major mergers (solid dots) and without (open circles) at $z=2$.
We see that the stellar distribution within $r_e$ typically resembles 
an oblate spheroid with the half-height approximately equal to one half of
that of the disk radius or more.
For such hot stellar systems no barlike equilibria exist and no strong
stellar bars would form secularly \citep[e.g.,][]{1973Ostriker}.
Indeed, we do not find any instance of thin stellar bars that would occupy locations 
near the left y-axis;
the one instance seen is in fact a close merging pair, which, when approximated as an ellipsoid by our code,
shows up as a thin bar.
The right panel of Figure~\ref{fig:DeDe} shows the same for SFR density,
which shows that ongoing star formation in the central region for the majority of galaxies at $z\ge 1$ 
takes place on a relatively thin disk of typical height-to-radius ratio of $0.1-0.3$,
with some ratios reaching as low as $0.03$, approaching our resolution limit of $\sim 100$pc.
It is clear, however, the relatively thick stellar bulges seen in the 
left panel of Figure~\ref{fig:DeDe} are very well resolved and little affected by resolution effects.
The number of stellar particles within ${\rm r_e}$ for mass in the range $10^9-10^{12}\msun$ 
are typically $N\sim 10^{3.5}-10^{6.5}$ and the two-body relaxation time 
is roughly $t_{r}\approx (N/50)t_c$ \citep[e.g.,][]{1997Steinmetz}, where $t_c$ is the orbital period  at ${\rm r_e}$.
For a galaxy with $M_{\rm star}=10^{10}\msun$ ($N\sim 10^{4.2}$ within ${\rm r_e}$) 
and ${\rm r_e}\sim 0.5$kpc (see Figure~\ref{fig:reMstarz2} below)
the relaxation heating time is estimated to be $\sim 1\times 10^{10}$yr.
A typical galaxy with $M_{\rm star}=10^{10}\msun$ corresponds to ${\rm SFR}\sim 10\msun$/yr at
the relevant redshift range.
Thus, we expect the two-body relaxation heating to be completely negligible 
for galaxies with ${\rm SFR}\ge 10\msun$/yr.
This shows that the dynamically hot state of the central stellar bulges of the simulated galaxies
is unlikely caused by numerical effects.
In the left panel of Figure~\ref{fig:DeDe} 
we do not see significant difference between galaxies with major mergers and those without,
indicating that major mergers do not appear to enhance formation of structures that resemble bars;
this issue will be further examined below.

\begin{figure}[ht]
\centering
\vskip -0cm
\resizebox{3.2in}{!}{\includegraphics[angle=0]{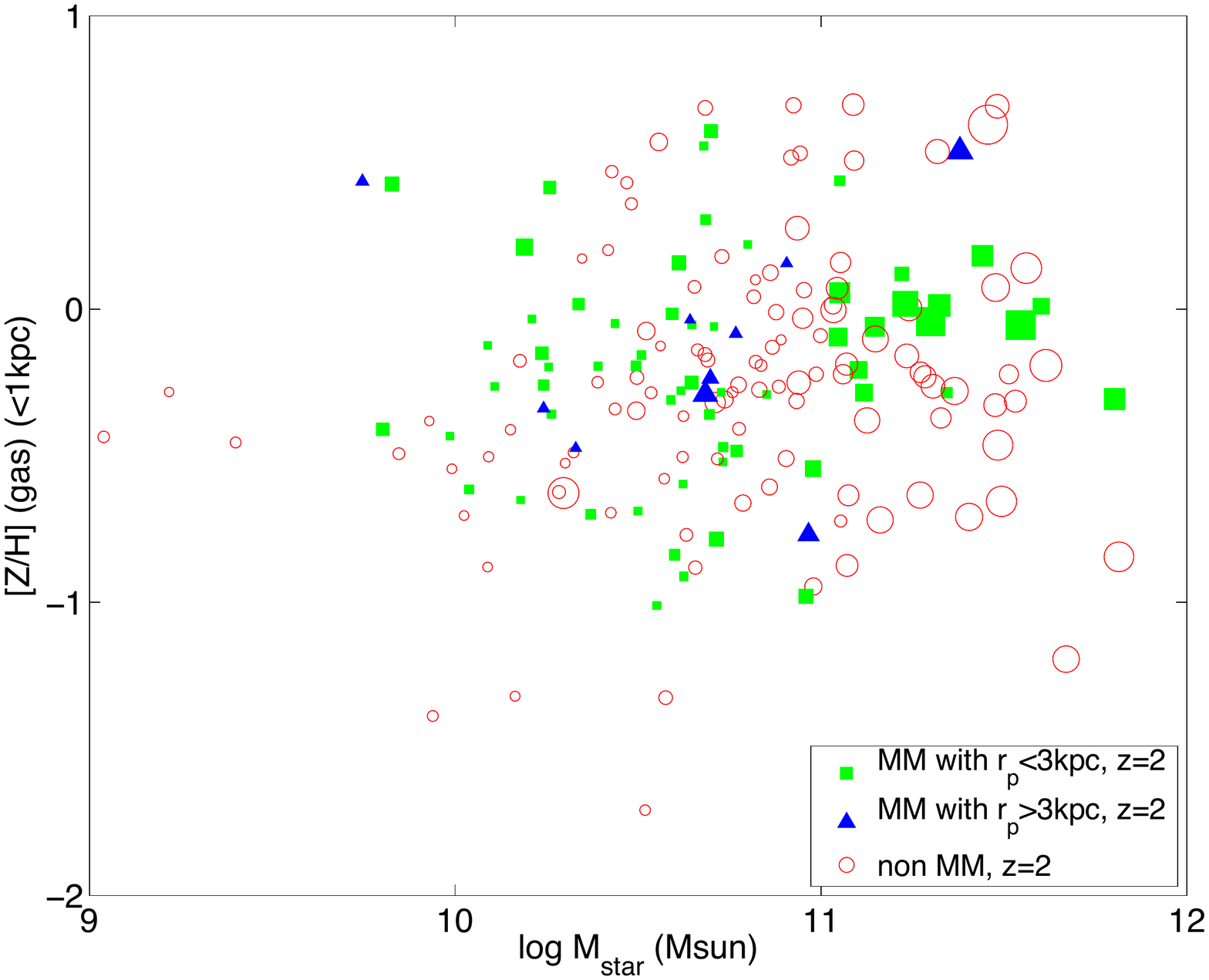}}
\resizebox{3.2in}{!}{\includegraphics[angle=0]{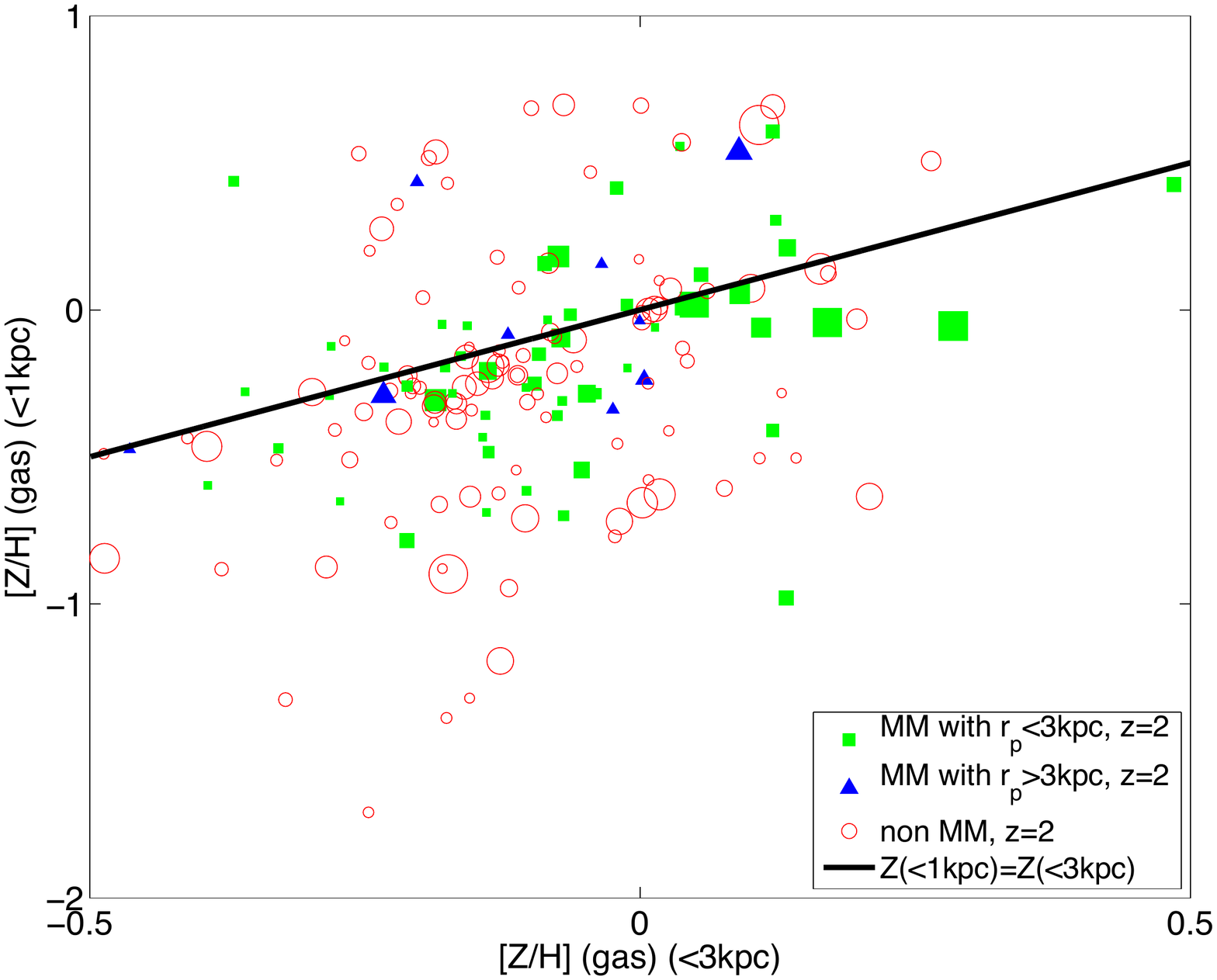}}
\vskip -0.4cm
\caption{\footnotesize 
Left panel: the mean gas metallicity within the central 1kpc region as a function of galaxy stellar mass
for galaxies with ${\rm SFR}\ge 10\msun$yr$^{-1}$.
Three types of galaxies are shown using different symbols: galaxies that are not undergoing major mergers are 
open circles, galaxies in major mergers with projected separation between the two galaxies
less than $3$kpc as squares and greater than $3$kpc as triangles.
Right panel: the mean gas metallicity within the central 1kpc region as a function of 
the mean gas metallicity within the central 3kpc region galaxy stellar mass
for galaxies with ${\rm SFR}\ge 10\msun$yr$^{-1}$.
The symbol size in both panels is linearly proportional to the logrithm of its SFR.
}
\label{fig:MetalMstar}
\end{figure}

In the absence of strong stellar bars, can significant gas inflows still exist?
Figure~\ref{fig:tgasdepletion} 
shows the gas depletion time in central regions at $r<1$kpc (left panel) 
and over the entire galaxy within the virial radius (right panel)
for all galaxies with ${\rm SFR}\ge 10\msun$yr$^{-1}$.
The right panel indicates that the gas depletion time over the entire galaxy is longer than
its dynamic time and comparable to the Hubble time.
The gas depletion time in the central $1$kpc region, however, is shorter at $\le 100$Myrs.
The gas depletion time in the central region spans a wide range, $0.1-100$Myrs,
and there is no discernible difference between galaxies in major mergers (solid symbols) and those that are not (open symbols).
Furthermore, there is no visible dependence of the depletion time in the central region 
on the separation of the two merging galaxies for those that are in major mergers.
Examination of SF histories of individual galaxies indicate that 
the SFR are relatively steady and their durations are on the order of Hubble time,
i.e., much longer than the gas depletion time scales of the central regions 
but comparable to the gas depletion time scales within the virial radii shown in Figure~\ref{fig:tgasdepletion}
(Figure~\ref{fig:majorSFR} shows that for galaxies with mergers within $\Delta z=0.5$).
This suggests that, irrespective of being in major mergers or not, gas inflows to the central regions 
appear to be ubiquitous; in other words, galaxies that are not in major mergers appear to be able to 
channel a sufficient amount of gas to fuel the star formation
on time scales that are much longer than the gas depletion times in the central regions.
The disparity in the gas depletion time scales of the central regions 
and between those and the overall star formation durations 
strongly imply that gas inflows, in general, are not smooth but in the form 
of clumps falling in intermittently.

To further demonstrate that gas inflows towards the central regions are generally not caused
by central non-spherical gravitational perturbations,
the left panel of Figure~\ref{fig:MetalMstar} 
shows the mean gas metallicity in the central $1$kpc region
and the right panel shows 
the mean gas metallicity in the central $1$kpc region as a function of 
the mean gas metallicity in the central $3$kpc region, comparing galaxies with and without major mergers.
From both panels we see that 
there is no visible difference in the metallicity of gas in the central regions
between galaxies that are in major mergers and those that are not.
It is seen that there is a relatively large span of mean gas metallicity in the central $1$kpc region,
from $\sim -1.5$ to $\sim 0.5$ for both types of galaxies,
while the range shrinks to about $-0.5$ to $0.5$ within $3$kpc for both types of galaxies.
If non-spherical gravitational perturbations in the central regions were responsible for driving
gas inward, they would be most effective for the gas in the immediate neighborhood.
Consequently, if the central $1$kpc region were just fed by gas driven inward from the immediate surroundings 
by internal non-spherical gravitational perturbations within,
one would expect to see a higher gas metallicity in the central $1$kpc than in the central $3$kpc,
since star formation rate is super-linear on gas density (the Schmidt-Kennicutt law) hence
SFR density stronger in the 1kpc central region than in the 3kpc central region per unit gas.
This expectation is not universally borne out for all galaxies in the simulation;
on the contrary, the majority of galaxies lie below the $Z(<1{\rm kpc})=Z(<3{\rm kpc})$ line, 
and there exists low mean metallicity ($Z<-0.5$) gas in the central $1$kpc that is not seen in the mean metallicity 
within the central $3$kpc. 
This is unambiguous evidence that a significant amount of gas inflow is directly ``parachuted in" 
(e.g., dynamical friction inspiral of gas clumps with or without dark matter halos, or infalling satellites on nearly radial orbits)
or ``channelled in" (e.g., clumpy cold streams)
from large scales, not smooth gas from regions that immediately surround it, at least for a large fraction of galaxies.
This is consistent with the implied intermittency of fueling seen in 
Figure~\ref{fig:tgasdepletion}.
In any event, the results indicate that major mergers do not appear to form a distinct set 
of galaxies with respective to gas metallicity in the central regions.

\begin{figure}[ht]
\centering
\vskip -0cm
\hskip -0.7cm
\resizebox{3.55in}{!}{\includegraphics[angle=0]{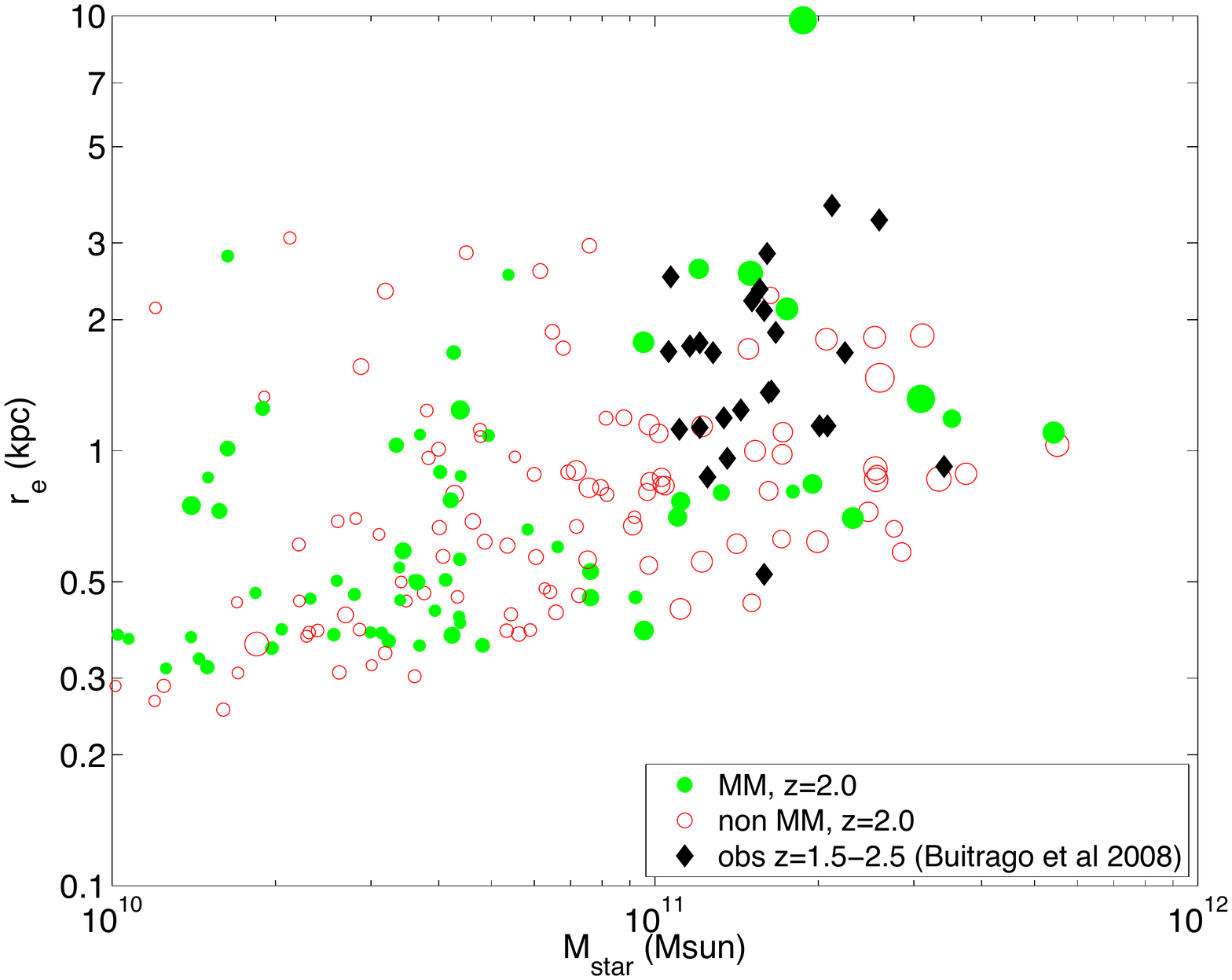}}
\hskip -1.0cm
\resizebox{3.55in}{!}{\includegraphics[angle=0]{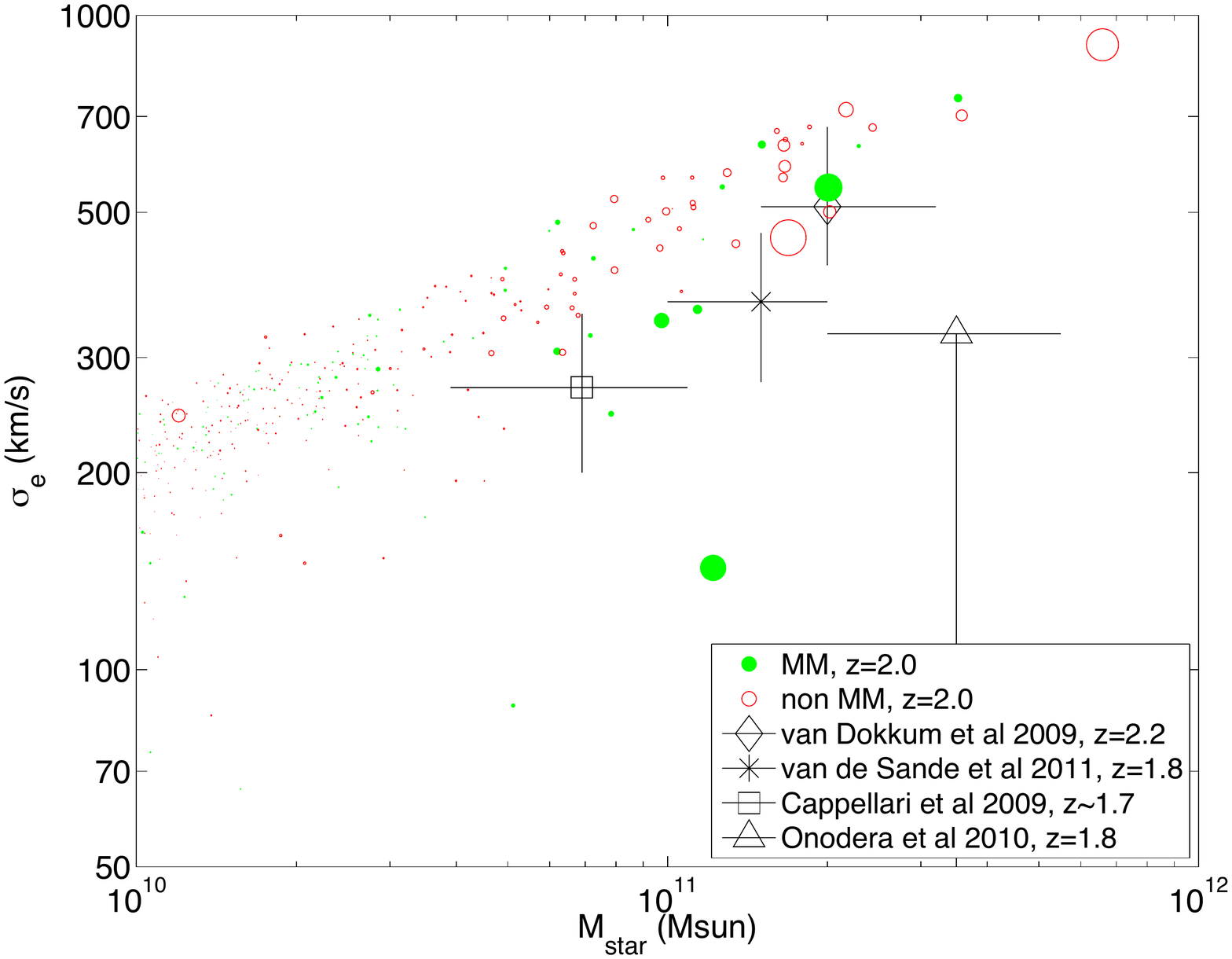}}
\vskip 0cm
\caption{\footnotesize 
Left panel: the effective radii of galaxies in restframe V band (observed H band) 
versus the stellar masses
for galaxies with major mergers (solid dots) and those without (open circles) at $z=2$.
Also shown as solid diamonds are the observations of 
\citet[][]{2008Buitrago} for the subset of galaxies at $z=1.5-2.5$ observed in H band.
The symbol size in the left panel is linearly proportional to the logrithm of SFR.
Right panel: the relation between velocity dispersion (y axis) and dynamical mass (x axis) at $r_e$.
The black diamond, star, square and triangle symbols with cross errorbars 
are the observational data for galaxies in the range range $z=17-2.2$ 
from \citet[][]{2009vanDokkum}, 
\citet[][]{2011vandeSande},
\citet[][]{2009Cappellari}
and 
\citet[][]{2010Onodera}, respectively.
The symbol size in the right panel is linearly proportional to the SFR.
}
\label{fig:reMstarz2}
\end{figure}

\citet[][]{1996Mihos} show that
galaxy structure plays a dominant role in regulating gas inflows, which they find 
are generally driven by gravitational torques from the host galaxy, rather than the companion, 
in their major merger simulations.
The lack of any significant merger induced effects appear at odds with their simulations at first instance.
We attribute the difference primarily to the difference in the physical properties of galaxies between 
merger simulations and those in present cosmological simulation at $z>1$.
Specifically, as we will show shortly, most of galaxies in our simulation appear to have massive stellar bulges,
whereas merger simulations with dramatic inflows seen during mergers start with pre-merger disk galaxies 
without massive stellar bulges.
In fact,  a subset of simulations by \citet[][]{1996Mihos} where the per-merger galaxies
have massive stellar bulges has already provided
insight for the above apparent discrepancy: 
they note that dense bulges act to stabilize galaxies against bar modes and have much diminished inflow enhancement.

In the left panel of Figure~\ref{fig:reMstarz2} we show the effective stellar radii
in restframe V band of galaxies with ${\rm SFR}\ge 10\msun$yr$^{-1}$ at $z=2$,
compared to observed galaxies also in restframe V band (observed H band).
We see that the effective radii of most simulated galaxies 
at $z=2$ are in the range of $0.5-2$kpc for galaxies of stellar mass $\ge 10^{11}\msun$,
consistent with previous results \citep[][]{2009Joung, 2009Naab}, and are in reasonable agreement with observations.
No dust obscuration is applied in the calculation so the computed radii are likely lower limits;
if we had taken dust obscuration into account, we expect the agreement would still be better.
The right panel of Figure~\ref{fig:reMstarz2} shows the 1-d velocity dispersion at the effective stellar radius
as a a function of stellar mass and we find that within the uncertainties the simulation results are in agreement with 
the observations, indicative of a self-consistency of the simulation results.
The observed high value of central velocity dispersion \citep[][]{2009vanDokkum} was somewhat surprising initially
based on an extrapolation of local elliptical galaxy properties, but now that 
additional observations have confirmed the earlier discovery and our simulations 
indicate that this is in fact in line with the theoretical expectation based on the cold dark matter model.
There is one exception \citep[][]{2010Onodera} that shows a lower central velocity dispersion;
our current statistics are insufficient to gauge this against our model one way or another.
Although the simulation results and observations are statistically consistent with one another,
enlarging both the simulation size and observed galaxy sample size may provide very useful 
constraints on physical processes that govern the formation of the bulges.
If pressed, one might incline to conclude that there is a slight hint that the simulated galaxies are 
slightly smaller than the handful of observed galaxies,
although the observed ones overlap and are statistically consistent with the simulated range in terms of velocity 
dispersion at a fixed stellar mass.
Nonetheless, three effects may have caused slight overestimation of the velocity dispersions of simulated galaxies.
First, no dust obscuration effect is taken into account.
Second, no observational beam smearing effect is taken into account.
Third, the simulated galaxies at a fixed stellar mass have a range in SFR,
whereas the observed galaxies shown are thought to be quiescent; 
one might think that gas loss from aging or dying stars acts in the direction of enlarging stellar cores with 
aging stellar population due to adiabatic expansion related to mass loss that is known to be substantial.

To be prudent and conservative, we have purposely plotted the symbol size in 
the right panel of Figure~\ref{fig:reMstarz2} to be linearly proportional to the SFR to see
if there is noticeable trend in SFR with core size/velocity dispersion.
We see one case, the green solid dot at ($1.3\times 10^{11}\msun$,150km/s), that 
has a SFR that may be a factor of a few higher than typical galaxies at around that mass.
However, we also see galaxies to have higher SFR even though having much higher velocity dispersions,
with or without major mergers.
In any case, it appears that some of the noticeably high SFR galaxies 
are consistent with being randomly distributed with respect to $\sigma_e$.
Thus, we conclude that there is no dramatic trend of SFR with respect 
to  $\sigma_e$ at a fixed stellar mass in the range of $\sigma_e$ that overlaps with observed values,
save the one noted exceptions that is presently difficult to gauge statistically.
This check suggests that our results are not hinged on our modeling of 
the size of the central stellar bulges being perfectly correct
and are thus robust to possible small variations.
Taking the evidence presented in the preceding four figures together a consistent physical picture emerges:

$\bullet$ Gravitational or hydrodynamic torques stemming from scales larger than 
the central regions containing most of the stars in the primary galaxy may play a fundamental role 
in transporting the necessary amount of gas to fuel the star formation in the central regions.

$\bullet$ A large portion of often metal-poor gas from large scales is directly transported into the central regions,
possibly in the form of dynamical friction inspiraling gas clumps, infalling satellites on nearly radial orbits,
or clumpy cold streams from large scales in an intermittent fashion.

$\bullet$ Significant gas inflows, not necessarily requiring major mergers,
allow for formation of dense, compact, not-so-flat stellar bulges that are stable to bar formation.

$\bullet$ Major mergers of galaxies, most of which have dense bulges, do not dramatically enhance 
gas inflows and SFR or cause significant differences in gas properties in the central regions
for galaxies at $z\ge 1$,
in accord with earlier major mergers simulations of disk galaxies with massive bulges.

\section{Conclusions}

With high resolution and a physically sound treatment of relevant physical processes,
our state-of-the-art, adaptive mesh-refinement Eulerian cosmological hydrodynamic simulations 
have reproduced well some key observables of the galaxy population as a whole \citep[][]{2010Cen, 2011bCen, 2011cCen},
including galaxy luminosity functions 
at $z=0-3$, galaxy color distribution at $z=0$, the entire star formation history, 
and properties of damped Lyman alpha systems that we have so far examined.
Here we study how major mergers affect star formation.
The simulation contains about $2000-3000$ galaxies 
with stellar masses in the range $10^9-10^{12}\msun$ and resolved at better than $114h^{-1}$pc at $z=1.4-3.7$,
providing a good statistical sample to examine major mergers for a wide of range of galaxies in mass and SFR.

The most significant finding is that major mergers, on average, do not 
result in two orders of magnitude boost in SFR, as found in simulations
of major mergers of gas-rich disk galaxies with idealized initial conditions.
Rather, for the redshift range examined, $z=1.4-3.7$, 
major mergers give rise to an average boost $0-60\%$ in specific SFR for SFR in range of $1-1000\msun$/yr examined.
Two physical factors of cosmological origin that are not taken into account in isolated merger simulations
may be responsible for the difference.

First, the central regions ($\sim 1$kpc) of galaxies at $z>1$, in the absence of major mergers,
are being fed, in an intermittent fashion,
with significant gas inflows.
As a result, galaxies without major mergers at $z>1$ have much higher SFR 
than their lower redshift counterparts, a fact that is known observationally.
We demonstrate that, at least for a significant fraction of galaxies,
gas inflows to the central regions, often quite metal poor,
originate from large scales (not smooth gas from
the regions immediately surrounding the central region)
possibly in the form of dynamical friction inspiraling gas clumps, infalling satellites on nearly radial orbits,
or clumpy cold streams from large scales.
We suggest that gravitational or hydrodynamic torques stemming from scales larger than 
the central regions play a fundamental role 
in transporting the necessary amount of gas to fuel the star formation in the central regions.
How this is achieved physically and which processes are most important 
are some of the very important issues that will be investigated in a future study.

Second, the large inflows of gas in galaxies with or without major mergers
produce compact, dense stellar cores/bulges with high velocity dispersions that 
are in agreement with observations and stable to bar formation.
The dense massive stellar bulges significantly diminish the importance 
of the major mergers induced, additional gas inflows for galaxies at $z\ge 1$,
in good agreement with earlier major mergers simulations of disk galaxies with massive bulges.

This result implies that a substantial revision of the current theoretical framework for 
galaxy formation is necessary, since some of the major foundational elements in our interpretation
of galaxy properties hinge on the requirements/beliefs that major mergers 
are responsible for some of the extreme galaxy formation events, 
including high luminosity galaxies, such as starbursting galaxies, ULIRGs and SMGs, and 
formation of supermassive black holes.

Some additional results found that may also be interesting are:

$\bullet$ $10-20\%$ of galaxies with stellar mass greater than $10^{11}\msun$ are in major mergers
at any time from $z=1-4$.

$\bullet$ The merger rate per unit redshift is roughly constant at $\sim 0.4$
for galaxies in the stellar mass range of $10^{10.7}-10^{11.7}\msun$ with an upturn
and a dramatic downturn above and below that mass range, respectively,
for the redshift range $z\sim 1-4$.
A fitting formula is provided in Eq~\ref{eq:R}.

$\bullet$ For galaxies with SFR greater than $200\msun$/yr we predict that about $10-40\%$ should be seen
in major mergers at $z=1-4$. This predicted fraction is somewhat
lower than what current spectral observations suggest \citep[e.g., 57\%;][]{2006Tacconi}
but can be directly tested with high resolution imaging with ALMA.

$\bullet$ It is predicted that the cumulative probability distribution function of major merging galaxies 
within a projected separation ${\rm r_p}$  
goes approximately as ${\rm r_p}^{1/4}$ for galaxies with SFR$\ge 10\msun$/yr (for ${\rm r_p}$ greater than a few kpc).
We expect that ALMA may be able to provide a direct measurement to test this.

\vskip 1cm

Computing resources were in part provided by the NASA High-
End Computing (HEC) Program through the NASA Advanced
Supercomputing (NAS) Division at Ames Research Center.
This work is supported in part by grants NAS8-03060 and NNX11AI23G.


\begin{thebibliography}{42}
\expandafter\ifx\csname natexlab\endcsname\relax\def\natexlab#1{#1}\fi

\bibitem[{{Abel} {et~al.}(1997){Abel}, {Anninos}, {Zhang}, \&
  {Norman}}]{1997Abel}
{Abel}, T., {Anninos}, P., {Zhang}, Y., \& {Norman}, M.~L. 1997, New Astronomy,
  2, 181

\bibitem[{{Bahcall} {et~al.}(1997){Bahcall}, {Kirhakos}, {Saxe}, \&
  {Schneider}}]{1997Bahcall}
{Bahcall}, J.~N., {Kirhakos}, S., {Saxe}, D.~H., \& {Schneider}, D.~P. 1997,
  \apj, 479, 642

\bibitem[{{Barnes} \& {Hernquist}(1996)}]{1996Barnes}
{Barnes}, J.~E., \& {Hernquist}, L. 1996, \apj, 471, 115

\bibitem[{{Bruzual} \& {Charlot}(2003)}]{Bruzual03}
{Bruzual}, G., \& {Charlot}, S. 2003, \mnras, 344, 1000

\bibitem[{Bryan \& Norman(1999)}]{1999bBryan}
Bryan, G.~L., \& Norman, M.~L. 1999, in Structured Adaptive Mesh Refinement
  Grid Methods, ed. N.~P.~C. S.~B.~Baden (IMA Volumes on Structured Adaptive
  Mesh Refinement Methods, No. 117), 165

\bibitem[{{Buitrago} {et~al.}(2008){Buitrago}, {Trujillo}, {Conselice},
  {Bouwens}, {Dickinson}, \& {Yan}}]{2008Buitrago}
{Buitrago}, F., {Trujillo}, I., {Conselice}, C.~J., {Bouwens}, R.~J.,
  {Dickinson}, M., \& {Yan}, H. 2008, \apjl, 687, L61

\bibitem[{{Cappellari} {et~al.}(2009){Cappellari}, {di Serego Alighieri},
  {Cimatti}, {Daddi}, {Renzini}, {Kurk}, {Cassata}, {Dickinson},
  {Franceschini}, {Mignoli}, {Pozzetti}, {Rodighiero}, {Rosati}, \&
  {Zamorani}}]{2009Cappellari}
{Cappellari}, M., {di Serego Alighieri}, S., {Cimatti}, A., {Daddi}, E.,
  {Renzini}, A., {Kurk}, J.~D., {Cassata}, P., {Dickinson}, M., {Franceschini},
  A., {Mignoli}, M., {Pozzetti}, L., {Rodighiero}, G., {Rosati}, P., \&
  {Zamorani}, G. 2009, \apjl, 704, L34

\bibitem[{{Cen}(2010)}]{2010Cen}
{Cen}, R. 2010, ArXiv e-prints

\bibitem[{{Cen}(2011{\natexlab{a}})}]{2011bCen}
---. 2011{\natexlab{a}}, ApJ, in press, arXiv1104.5046

\bibitem[{{Cen}(2011{\natexlab{b}})}]{2011cCen}
---. 2011{\natexlab{b}}, ArXiv e-prints

\bibitem[{{Cen} {et~al.}(1995){Cen}, {Kang}, {Ostriker}, \& {Ryu}}]{1995Cen}
{Cen}, R., {Kang}, H., {Ostriker}, J.~P., \& {Ryu}, D. 1995, \apj, 451, 436

\bibitem[{{Cen} {et~al.}(2005){Cen}, {Nagamine}, \& {Ostriker}}]{2005Cen}
{Cen}, R., {Nagamine}, K., \& {Ostriker}, J.~P. 2005, \apj, 635, 86

\bibitem[{{Cen} \& {Ostriker}(1992)}]{1992CenOstriker}
{Cen}, R., \& {Ostriker}, J.~P. 1992, \apjl, 399, L113

\bibitem[{{Cimatti} {et~al.}(2008){Cimatti}, {Cassata}, {Pozzetti}, {Kurk},
  {Mignoli}, {Renzini}, {Daddi}, {Bolzonella}, {Brusa}, {Rodighiero},
  {Dickinson}, {Franceschini}, {Zamorani}, {Berta}, {Rosati}, \&
  {Halliday}}]{2008Cimatti}
{Cimatti}, A., {Cassata}, P., {Pozzetti}, L., {Kurk}, J., {Mignoli}, M.,
  {Renzini}, A., {Daddi}, E., {Bolzonella}, M., {Brusa}, M., {Rodighiero}, G.,
  {Dickinson}, M., {Franceschini}, A., {Zamorani}, G., {Berta}, S., {Rosati},
  P., \& {Halliday}, C. 2008, \aap, 482, 21

\bibitem[{{Daddi} {et~al.}(2005){Daddi}, {Renzini}, {Pirzkal}, {Cimatti},
  {Malhotra}, {Stiavelli}, {Xu}, {Pasquali}, {Rhoads}, {Brusa}, {di Serego
  Alighieri}, {Ferguson}, {Koekemoer}, {Moustakas}, {Panagia}, \&
  {Windhorst}}]{2005Daddi}
{Daddi}, E., {Renzini}, A., {Pirzkal}, N., {Cimatti}, A., {Malhotra}, S.,
  {Stiavelli}, M., {Xu}, C., {Pasquali}, A., {Rhoads}, J.~E., {Brusa}, M., {di
  Serego Alighieri}, S., {Ferguson}, H.~C., {Koekemoer}, A.~M., {Moustakas},
  L.~A., {Panagia}, N., \& {Windhorst}, R.~A. 2005, \apj, 626, 680

\bibitem[{{Dalgarno} \& {McCray}(1972)}]{1972Dalgarno}
{Dalgarno}, A., \& {McCray}, R.~A. 1972, \araa, 10, 375

\bibitem[{{Di Matteo} {et~al.}(2005){Di Matteo}, {Springel}, \&
  {Hernquist}}]{2005DiMatteo}
{Di Matteo}, T., {Springel}, V., \& {Hernquist}, L. 2005, \nat, 433, 604

\bibitem[{{Duc} {et~al.}(1997){Duc}, {Mirabel}, \& {Maza}}]{1997Duc}
{Duc}, P.-A., {Mirabel}, I.~F., \& {Maza}, J. 1997, \aaps, 124, 533

\bibitem[{Eisenstein \& Hu(1999)}]{1999Eisenstein}
Eisenstein, D., \& Hu, P. 1999, ApJ, 511, 5

\bibitem[{{Haardt} \& {Madau}(1996)}]{1996Haardt}
{Haardt}, F., \& {Madau}, P. 1996, \apj, 461, 20

\bibitem[{{Heckman}(2001)}]{2001Heckman}
{Heckman}, T.~M. 2001, in Astronomical Society of the Pacific Conference
  Series, Vol. 240, Gas and Galaxy Evolution, ed. J.~E. {Hibbard}, M.~{Rupen},
  \& J.~H. {van Gorkom}, 345

\bibitem[{{Hopkins} {et~al.}(2006){Hopkins}, {Hernquist}, {Cox}, {Di Matteo},
  {Robertson}, \& {Springel}}]{2006Hopkins}
{Hopkins}, P.~F., {Hernquist}, L., {Cox}, T.~J., {Di Matteo}, T., {Robertson},
  B., \& {Springel}, V. 2006, \apjs, 163, 1

\bibitem[{{Joseph} \& {Wright}(1985)}]{1985Joseph}
{Joseph}, R.~D., \& {Wright}, G.~S. 1985, \mnras, 214, 87

\bibitem[{{Joung} {et~al.}(2009){Joung}, {Cen}, \& {Bryan}}]{2009Joung}
{Joung}, M.~R., {Cen}, R., \& {Bryan}, G.~L. 2009, \apjl, 692, L1

\bibitem[{{Komatsu} {et~al.}(2010){Komatsu}, {Smith}, {Dunkley}, {Bennett},
  {Gold}, {Hinshaw}, {Jarosik}, {Larson}, {Nolta}, {Page}, {Spergel},
  {Halpern}, {Hill}, {Kogut}, {Limon}, {Meyer}, {Odegard}, {Tucker}, {Weiland},
  {Wollack}, \& {Wright}}]{2010Komatsu}
{Komatsu}, E., {Smith}, K.~M., {Dunkley}, J., {Bennett}, C.~L., {Gold}, B.,
  {Hinshaw}, G., {Jarosik}, N., {Larson}, D., {Nolta}, M.~R., {Page}, L.,
  {Spergel}, D.~N., {Halpern}, M., {Hill}, R.~S., {Kogut}, A., {Limon}, M.,
  {Meyer}, S.~S., {Odegard}, N., {Tucker}, G.~S., {Weiland}, J.~L., {Wollack},
  E., \& {Wright}, E.~L. 2010, ArXiv e-prints

\bibitem[{{Longhetti} {et~al.}(2007){Longhetti}, {Saracco}, {Severgnini},
  {Della Ceca}, {Mannucci}, {Bender}, {Drory}, {Feulner}, \&
  {Hopp}}]{2007Longhetti}
{Longhetti}, M., {Saracco}, P., {Severgnini}, P., {Della Ceca}, R., {Mannucci},
  F., {Bender}, R., {Drory}, N., {Feulner}, G., \& {Hopp}, U. 2007, \mnras,
  374, 614

\bibitem[{Lowenthal {et~al.}(1997)Lowenthal, Koo, Guzman, Gallego, Phillips,
  Faber, Vogt, Illingworth, {et~al.}}]{1997Lowenthal}
Lowenthal, J.~D., Koo, D.~C., Guzman, R., Gallego, J., Phillips, A.~C., Faber,
  S.~M., Vogt, N.~P., Illingworth, G.~D., {et~al.} 1997, ApJ, 481, 673

\bibitem[{{Lutz} {et~al.}(1998){Lutz}, {Spoon}, {Rigopoulou}, {Moorwood}, \&
  {Genzel}}]{1998Lutz}
{Lutz}, D., {Spoon}, H.~W.~W., {Rigopoulou}, D., {Moorwood}, A.~F.~M., \&
  {Genzel}, R. 1998, \apjl, 505, L103

\bibitem[{{Mihos} \& {Hernquist}(1996)}]{1996Mihos}
{Mihos}, J.~C., \& {Hernquist}, L. 1996, \apj, 464, 641

\bibitem[{{Naab} {et~al.}(2009){Naab}, {Johansson}, \& {Ostriker}}]{2009Naab}
{Naab}, T., {Johansson}, P.~H., \& {Ostriker}, J.~P. 2009, \apjl, 699, L178

\bibitem[{{Onodera} {et~al.}(2010){Onodera}, {Daddi}, {Gobat}, {Cappellari},
  {Arimoto}, {Renzini}, {Yamada}, {McCracken}, {Mancini}, {Capak}, {Carollo},
  {Cimatti}, {Giavalisco}, {Ilbert}, {Kong}, {Lilly}, {Motohara}, {Ohta},
  {Sanders}, {Scoville}, {Tamura}, \& {Taniguchi}}]{2010Onodera}
{Onodera}, M., {Daddi}, E., {Gobat}, R., {Cappellari}, M., {Arimoto}, N.,
  {Renzini}, A., {Yamada}, Y., {McCracken}, H.~J., {Mancini}, C., {Capak}, P.,
  {Carollo}, M., {Cimatti}, A., {Giavalisco}, M., {Ilbert}, O., {Kong}, X.,
  {Lilly}, S., {Motohara}, K., {Ohta}, K., {Sanders}, D.~B., {Scoville}, N.,
  {Tamura}, N., \& {Taniguchi}, Y. 2010, \apjl, 715, L6

\bibitem[{{Ostriker} \& {Peebles}(1973)}]{1973Ostriker}
{Ostriker}, J.~P., \& {Peebles}, P.~J.~E. 1973, \apj, 186, 467

\bibitem[{{Sanders} {et~al.}(1988){Sanders}, {Soifer}, {Elias}, {Madore},
  {Matthews}, {Neugebauer}, \& {Scoville}}]{1988Sanders}
{Sanders}, D.~B., {Soifer}, B.~T., {Elias}, J.~H., {Madore}, B.~F., {Matthews},
  K., {Neugebauer}, G., \& {Scoville}, N.~Z. 1988, \apj, 325, 74

\bibitem[{Steidel {et~al.}(2003)Steidel, Adelberger, Adelberger, Shapley,
  Pettini, Dickinson, \& Giavalisco}]{2003Steidel}
Steidel, C.~C., Adelberger, K.~L., Adelberger, K.~L., Shapley, A.~E., Pettini,
  M., Dickinson, M., \& Giavalisco, M. 2003, ApJ, 592, 728

\bibitem[{{Steinmetz} \& {White}(1997)}]{1997Steinmetz}
{Steinmetz}, M., \& {White}, S.~D.~M. 1997, \mnras, 288, 545

\bibitem[{{Tacconi} {et~al.}(2006){Tacconi}, {Neri}, {Chapman}, {Genzel},
  {Smail}, {Ivison}, {Bertoldi}, {Blain}, {Cox}, {Greve}, \&
  {Omont}}]{2006Tacconi}
{Tacconi}, L.~J., {Neri}, R., {Chapman}, S.~C., {Genzel}, R., {Smail}, I.,
  {Ivison}, R.~J., {Bertoldi}, F., {Blain}, A., {Cox}, P., {Greve}, T., \&
  {Omont}, A. 2006, \apj, 640, 228

\bibitem[{{Toft} {et~al.}(2007){Toft}, {van Dokkum}, {Franx}, {Labbe},
  {F{\"o}rster Schreiber}, {Wuyts}, {Webb}, {Rudnick}, {Zirm}, {Kriek}, {van
  der Werf}, {Blakeslee}, {Illingworth}, {Rix}, {Papovich}, \&
  {Moorwood}}]{2007Toft}
{Toft}, S., {van Dokkum}, P., {Franx}, M., {Labbe}, I., {F{\"o}rster
  Schreiber}, N.~M., {Wuyts}, S., {Webb}, T., {Rudnick}, G., {Zirm}, A.,
  {Kriek}, M., {van der Werf}, P., {Blakeslee}, J.~P., {Illingworth}, G.,
  {Rix}, H., {Papovich}, C., \& {Moorwood}, A. 2007, \apj, 671, 285

\bibitem[{{Toomre} \& {Toomre}(1972)}]{1972Toomre}
{Toomre}, A., \& {Toomre}, J. 1972, \apj, 178, 623

\bibitem[{{Trujillo} {et~al.}(2006{\natexlab{a}}){Trujillo}, {Feulner},
  {Goranova}, {Hopp}, {Longhetti}, {Saracco}, {Bender}, {Braito}, {Della Ceca},
  {Drory}, {Mannucci}, \& {Severgnini}}]{2006bTrujillo}
{Trujillo}, I., {Feulner}, G., {Goranova}, Y., {Hopp}, U., {Longhetti}, M.,
  {Saracco}, P., {Bender}, R., {Braito}, V., {Della Ceca}, R., {Drory}, N.,
  {Mannucci}, F., \& {Severgnini}, P. 2006{\natexlab{a}}, \mnras, 373, L36

\bibitem[{{Trujillo} {et~al.}(2006{\natexlab{b}}){Trujillo}, {F{\"o}rster
  Schreiber}, {Rudnick}, {Barden}, {Franx}, {Rix}, {Caldwell}, {McIntosh},
  {Toft}, {H{\"a}ussler}, {Zirm}, {van Dokkum}, {Labb{\'e}}, {Moorwood},
  {R{\"o}ttgering}, {van der Wel}, {van der Werf}, \& {van
  Starkenburg}}]{2006Trujillo}
{Trujillo}, I., {F{\"o}rster Schreiber}, N.~M., {Rudnick}, G., {Barden}, M.,
  {Franx}, M., {Rix}, H., {Caldwell}, J.~A.~R., {McIntosh}, D.~H., {Toft}, S.,
  {H{\"a}ussler}, B., {Zirm}, A., {van Dokkum}, P.~G., {Labb{\'e}}, I.,
  {Moorwood}, A., {R{\"o}ttgering}, H., {van der Wel}, A., {van der Werf}, P.,
  \& {van Starkenburg}, L. 2006{\natexlab{b}}, \apj, 650, 18

\bibitem[{{van de Sande} {et~al.}(2011){van de Sande}, {Kriek}, {Franx}, {van
  Dokkum}, {Bezanson}, {Whitaker}, {Brammer}, {Labb{\'e}}, {Groot}, \&
  {Kaper}}]{2011vandeSande}
{van de Sande}, J., {Kriek}, M., {Franx}, M., {van Dokkum}, P.~G., {Bezanson},
  R., {Whitaker}, K.~E., {Brammer}, G., {Labb{\'e}}, I., {Groot}, P.~J., \&
  {Kaper}, L. 2011, \apjl, 736, L9+

\bibitem[{{van Dokkum} {et~al.}(2009){van Dokkum}, {Kriek}, \&
  {Franx}}]{2009vanDokkum}
{van Dokkum}, P.~G., {Kriek}, M., \& {Franx}, M. 2009, \nat, 460, 717

\end{thebibliography}

\end{document}